\documentclass[prl,superscriptaddress, noshowpacs,twocolumn]{revtex4}
\usepackage[dvips]{graphicx}
\usepackage{amssymb}

\newcommand{\ped}[1]{\ensuremath{_{\rm #1}}}

\begin{document}

\title{A Point-Contact Study of the Superconducting Gaps in
Al-Substituted and C-Substituted MgB$_2$ Single Crystals}

\author{R.\@S. Gonnelli}
\affiliation{Dipartimento di Fisica and INFM, Politecnico di
Torino, 10129 Torino, Italy}
\affiliation{INFM - LAMIA, 16152 Genova, Italy }%
\author{D. Daghero}
\affiliation{Dipartimento di Fisica and INFM, Politecnico di
Torino, 10129 Torino, Italy}
\affiliation{INFM - LAMIA, 16152
Genova, Italy }
\author{G.A. Ummarino}
\affiliation{Dipartimento di Fisica and INFM, Politecnico di
Torino, 10129 Torino, Italy}
\affiliation{INFM - LAMIA, 16152
Genova, Italy }
\author{A. Calzolari}
\affiliation{Dipartimento di Fisica and INFM, Politecnico di
Torino, 10129 Torino, Italy}
\author{Valeria Dellarocca}
\affiliation{Dipartimento di Fisica and INFM, Politecnico di
Torino, 10129 Torino, Italy}
\author{V.\@A. Stepanov}
\affiliation{P.N. Lebedev Physical Institute, Russian Academy of
Sciences, 119991 Moscow, Russia}
\author{S.M. Kazakov}
\affiliation{Solid State Physics Laboratory, ETH, CH-8093 Zurich,
Switzerland}
\author{J. Jun}
\affiliation{Solid State Physics Laboratory, ETH, CH-8093 Zurich,
Switzerland}
\author{J. Karpinski}
\affiliation{Solid State Physics Laboratory, ETH, CH-8093 Zurich,
Switzerland}

\begin{abstract}\vspace{1mm}
We present the results of directional point-contact spectroscopy in
state-of-the-art Mg$_{1-x}$Al$_x$B$_2$ and
Mg(B$\ped{1-y}$C$\ped{y}$)$\ped{2}$ single crystals produced at ETH,
Zurich. Fitting the conductance curves of our point contacts, that
always feature Andreev reflection structures, we obtained the doping
dependence of the gap amplitudes. The results are discussed in
comparison with other experimental findings and relevant theoretical
predictions. We conclude that the physics of Al-substituted crystals
at $x\gtrsim0.09$ might be governed by phase segregation, while
C-substituted crystals unexpectedly show a doping-induced transition
to single-gap superconductivity at $y=0.132$.
\end{abstract}
\maketitle Magnesium diboride, MgB$_2$, represents a unique and
lucky combination of different physical properties that make it
the highest-$T\ped{c}$ intermetallic compound, the only
superconducting diboride and the clearest example of two-band
superconductor ever discovered. As a matter of fact, most of its
physics has been explained rather well within the two-band model
in either the BCS~\cite{Liu} or the
Eliashberg~\cite{Brinkman,Choi} formulation . Much effort has been
made in order to understand whether the peculiar properties of
MgB$_2$ can be in some way tuned and controlled, both in view of
applications (where, for example, higher critical fields or
smaller anisotropy are required) or for fundamental reasons (to
test the predictions of the two-band models concerning the effects
of variations in some of the physical quantities that describe
MgB$_2$). In other words, most of the present research work is
devoted to investigate the ``neighborhood'' of MgB$_2$, that is
all the systems that can be obtained from MgB$_2$ by means of
pressure, irradiation, lattice stress, disorder and, over all,
chemical substitutions.

Obtaining partial substitution of Mg or B atoms in MgB$_2$ is a
difficult task. Even in the (few) cases of success, e.g. with
aluminum and carbon, there are problems of solubility
\cite{Bharathi}, phase segregation \cite{Barabash,Karpinski_Rio},
inhomogeneities\cite{Kazakov_C} and structural transitions
\cite{Slusky}. Most of the substituted samples presently available
are polycrystalline, of quality good enough to allow various kinds
of experimental investigations that have led to highlight many
structural, electronic and superconducting properties of these
compounds. However, only a few determinations of the doping
dependence of the energy gaps have appeared in literature, both
for Mg$_{1-x}$Al$_x$B$_2$ \cite{Putti,Putti_nuovo} and
Mg(B$\ped{1-y}$C$\ped{y}$)$\ped{2}$
\cite{Holanova,Schmidt,Papagelis}, and none in single crystals.

In the following, we will present the results of the first
systematic study of the energy gaps in Mg$_{1-x}$Al$_x$B$_2$ and
Mg(B$\ped{1-y}$C$\ped{y}$)$\ped{2}$ single crystals as a function
of Al and C contents, by means of \emph{directional} point-contact
spectroscopy (DPCS). We will show that the doping dependence of
the gaps is completely different in Al-substituted and
C-substituted samples. In Mg$_{1-x}$Al$_x$B$_2$ crystals there is
no evidence of gap merging and the small gap strongly decreases on
increasing $x$, to become as small as 0.4 meV at $x=0.21$, while
the large gap saturates at about 4 meV at high Al contents. In
Mg(B$\ped{1-y}$C$\ped{y}$)$\ped{2}$ crystals, instead, the
$\pi$-band gap remains practically unchanged (at most, it shows a
small increase), while the $\sigma$-band gap decreases and, at
$x=0.132$, the two gaps merge into one of amplitude $\Delta \simeq
3$~meV. The results will be discussed in comparison with other
experimental findings as well as with theoretical predictions.

\begin{figure}[b]
\vspace{-5mm}
\includegraphics[keepaspectratio,width=\columnwidth]{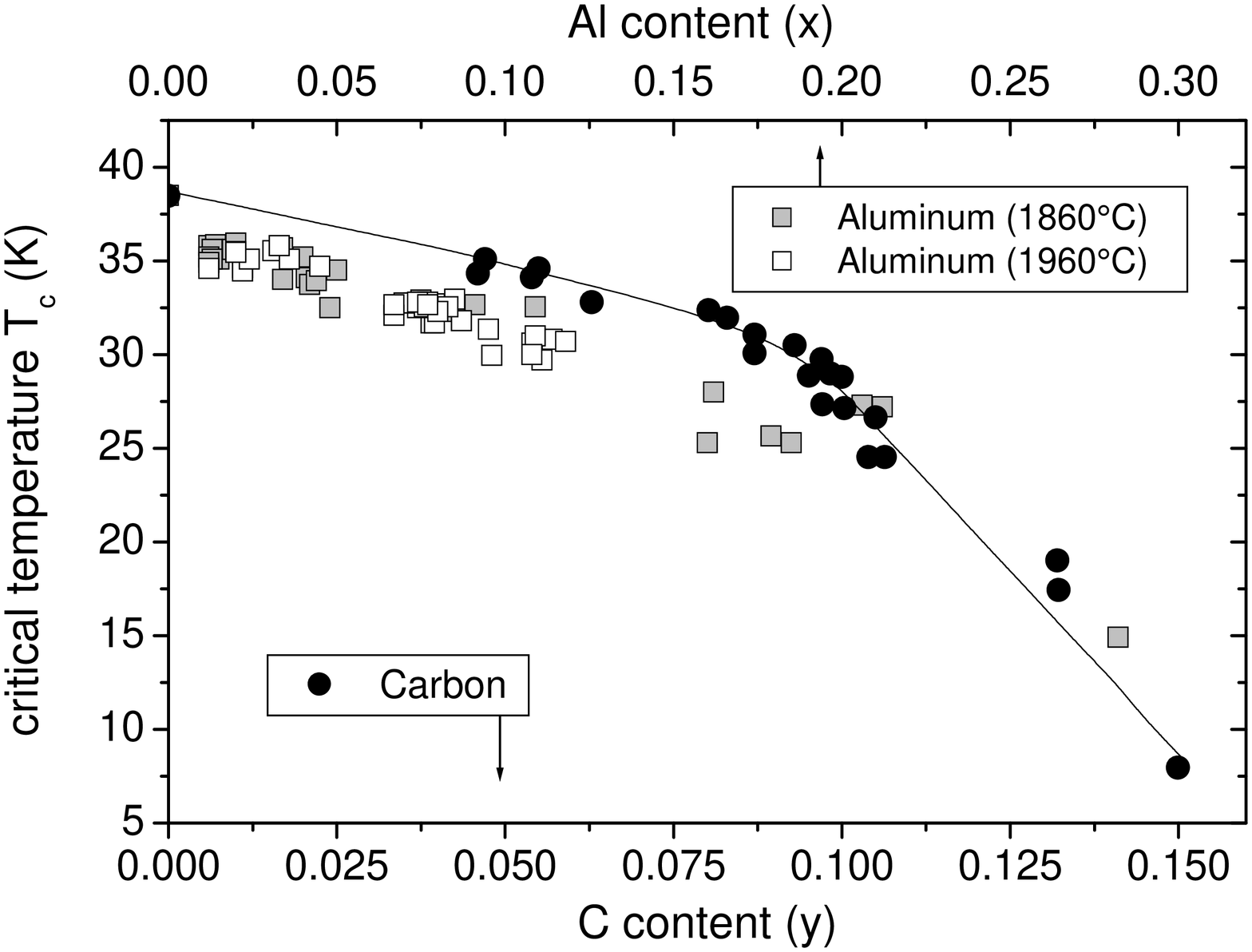}
\caption{Experimental dependence of the critical temperature
(evaluated from DC magnetization) on the content of Aluminum (top
axis) or Carbon (bottom axis) in Mg$_{1-x}$Al$_x$B$_2$ and
Mg(B$\ped{1-y}$C$\ped{y}$)$\ped{2}$, respectively. The two sets of
points for Mg$_{1-x}$Al$_x$B$_2$ refer to different growth
temperatures.}\label{fig:0}
\end{figure}

Both the Mg$_{1-x}$Al$_x$B$_2$ and the
Mg(B$\ped{1-y}$C$\ped{y}$)$\ped{2}$ single crystals were grown at
the Solid State Laboratory, ETH-Zurich (Switzerland) by using a
high-pressure technique in a cubic-anvil press, in the same way as
the unsubstituted crystals~\cite{crystal_growth}). The partial
substitution of Al in MgB$_2$ was obtained by replacing part of the
Mg precursor with Al \cite{Karpinski_Rio}. As evidenced by HRTEM, at
high doping levels there is a strong tendency to the precipitation
of a second phase, in the form of Al-rich layers (probably
MgAlB$_4$) perpendicular to the $c$ axis, while no defects are shown
in the $ab$ plane.
\begin{figure}[t]
\begin{center}
\includegraphics[keepaspectratio,width=0.7\columnwidth]{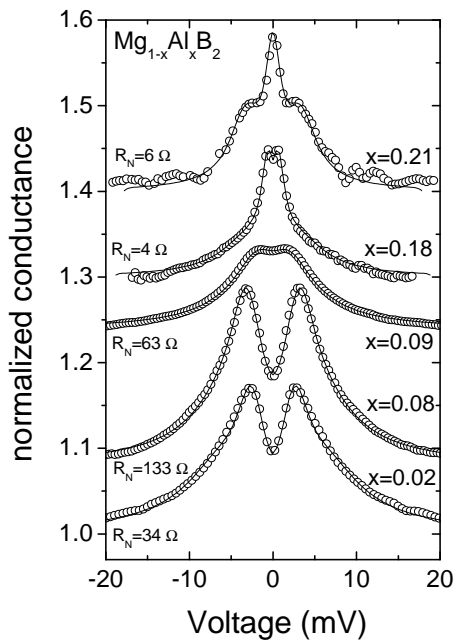}
\end{center}
\vspace{-5mm} \caption{Symbols: experimental normalized
conductance curves of Ag/Mg$_{1-x}$Al$_x$B$_2$ point contacts for
different Al contents $x$. The curves are vertically shifted for
clarity, and were all measured at 4.2 K apart from the last two
($x=0.18,\,021$) that were measured at 1.8 K. The normal-state
junction resistance is also indicated. Lines: best-fit curves
obtained with the two-band BTK model (see text for
details).}\label{fig:1}
\end{figure}
The Mg(B$\ped{1-y}$C$\ped{y}$)$\ped{2}$ crystals were grown at
1900-1950$^{\circ}$C by starting from magnesium, amorphous boron and
graphite powder or SiC as a carbon source. In the latter case, no
trace of Si was found in the final material \cite{Kazakov_C}.

The crystals used for our DPCS measurements had Al contents $x$
(measured with EDX) ranging from 0.02 up to 0.21, and C contents $y$
(evaluated from the cell parameter $a$) between 0.055 and 0.132.
Figure \ref{fig:0} reports the doping dependence of the critical
temperature $T\ped{c}$, given by DC magnetization measurements, for
both Mg$_{1-x}$Al$_x$B$_2$ and Mg(B$\ped{1-y}$C$\ped{y}$)$\ped{2}$
crystals. Very surprisingly, the two curves turn out to be rather
similar if plotted versus the atomic content of Al and C ($x$ and
$2y$, respectively).

Directional point-contact measurements were performed by using the
pressure-less (``soft'') technique described elsewhere
\cite{nostroPRL,nostroSUST} that consists in using a small
($\varnothing \leq 50 ~\mu$m) drop of Ag conductive paint as the
counterelectrode, instead of the usual metallic tip pressed against
the sample surface. This ensures greater contact stability on
thermal cycling and allows making the contacts on the side of the
(very thin) crystals, so as to inject the current mainly parallel to
the $ab$ planes. In unsubstituted MgB$_2$, this is the best
configuration for a contemporaneous measurement of \emph{both} the
gaps \cite{Brinkman,nostroPRL}. The experimental conductance curves
($\mathrm{d}I/\mathrm{d}V$ vs. $V$) of our point contacts were
normalized to the normal-state conductance to allow comparison with
the Blonder-Tinkham-Klapwijk (BTK) model for superconductor/normal
metal interfaces \cite{BTK}. All our contacts were in the ballistic
limit and had small potential barrier. Indeed, the conductance
curves show clear Andreev-reflection features. In particular, they
present clear maxima at energies roughly equal to the small gap,
$\Delta\ped{\pi}$, but (in spite of the current injection along the
$ab$ plane \cite{nostroPRL}) only smooth shoulders at energies
corresponding to the large gap, $\Delta\ped{\sigma}$.

\begin{figure}[t]
\begin{center}
\includegraphics[keepaspectratio,width=0.7\columnwidth]{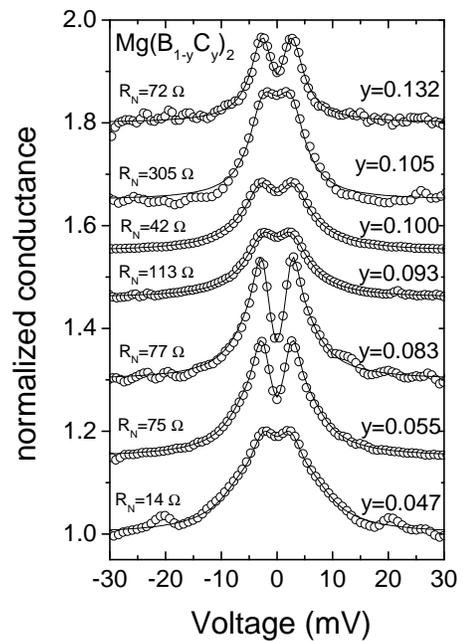}
\end{center}
\vspace{-5mm} \caption{Symbols: experimental normalized
conductance curves of Ag/Mg(B$\ped{1-y}$C$\ped{y}$)$\ped{2}$ point
contacts for different C contents $y$. The curves are vertically
shifted for clarity, and were all measured at 4.2 K. The
normal-state resistances are also indicated. Lines: best-fit
curves obtained with the two-band or one-band BTK
model.}\label{fig:2}
\end{figure}

Figure \ref{fig:1} reports some normalized experimental
conductance curves (symbols) measured in single crystals with
different Al contents. All the curves were recorded at 4.2 K apart
from the last two ($x=0.18,0.21$) that were measured at 1.8 K
because the thermal smearing at 4.2 K was already comparable to
the energy width of the Andreev-reflection structures. Even at a
first glance, two distinct doping regimes can be identified. For
$x< 0.09$, the conductance peaks corresponding to
$\Delta\ped{\pi}$ slightly move outwards with respect to the
undoped case. For $x>0.09$, instead, the peaks shrink very fast on
increasing $x$ and finally merge in a single sharp maximum at zero
bias, indicating a fast decrease in $\Delta\ped{\pi}$. The clear
narrowing of the whole conductance curves in passing from the
low-$x$ to the high-$x$ regime indicates that also
$\Delta\ped{\sigma}$ undergoes a significant (and sudden) change
around $x=0.09$.

A quantitative evaluation of the gap amplitudes can be given by
fitting the conductance curves with the BTK model generalized to the
two-band case, that has been shown to work well in pure
MgB$_2$~\cite{nostroPRL,nostroPRB}. The best-fitting curves, that
agree well with the experimental data, are shown as solid lines in
Fig.\ref{fig:1}. The fitting function contains 7 parameters: the
gaps $\Delta\ped{\sigma}$ and $\Delta\ped{\pi}$, the barrier
parameters $Z\ped{\sigma}$ and $Z\ped{\pi}$, the lifetime broadening
parameters $\Gamma\ped{\sigma}$ and $\Gamma\ped{\pi}$, plus the
weight of the $\pi$ band in the total conductance, $w\ped{\pi}$.
Hence, one could object that the fitting procedure should give
rather large uncertainties on the gap values. Actually: i) the value
of $\Delta\ped{\pi}$ is quite well determined by the energy position
of the conductance peaks, and thus its uncertainty is necessarily
small; ii) the values of both $\Delta\ped{\pi}$ and
$\Delta\ped{\sigma}$ were confirmed, up to $x=0.09$, by the
independent, three-parameter fit of the $\sigma$ and $\pi$-band
contributions to the conductance, whose separation was possible by
applying a suitable magnetic field to the junction, as explained
elsewhere \cite{nostroPRL,nostroPRB}. This was not possible for
$x\!\!>$0.09, where even weak fields depress the $\sigma$-band gap
and leave $\Delta\ped{\pi}$ almost unchanged.

Fig.\ref{fig:2} reports a subset of the conductance curves
measured at 4.2 K in various Mg(B$\ped{1-y}$C$\ped{y}$)$\ped{2}$
single crystals with different $y$. In this case there are no
dramatic changes in the amplitude of the small gap
$\Delta\ped{\pi}$, while the constant narrowing of the
Andreev-reflection features indicates a decrease in the
$\sigma$-band gap. As in Fig.\ref{fig:1}, solid lines represent
the two-band BTK best-fitting curves. In various cases, we were
able to separate (and fit separately) the partial $\sigma$ and
$\pi$-band conductances by applying a suitable magnetic field,
thus achieving a higher-precision determination of the gap
amplitudes. (Actually, the effect of the field on the conductance
curves of the Ag/Mg(B$\ped{1-y}$C$\ped{y}$)$\ped{2}$ point
contacts is rather complex and will be the subject of a
forthcoming paper).
\begin{figure}[t]
\vspace{-2mm}
\includegraphics[keepaspectratio,width=0.8\columnwidth]{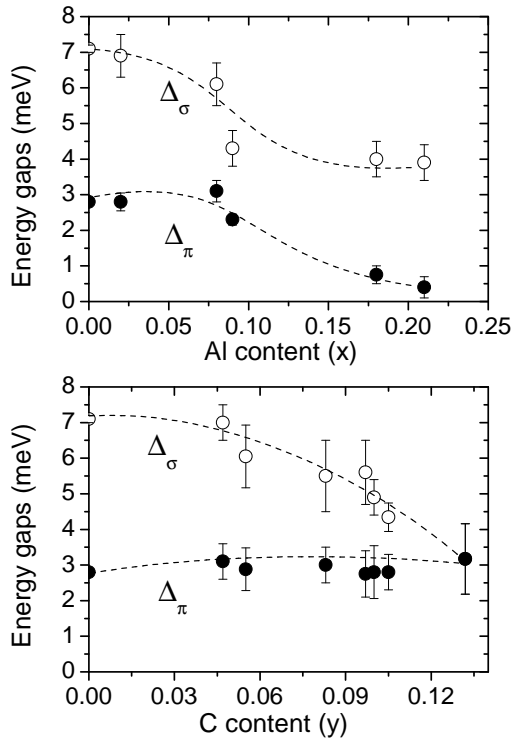}
\vspace{-2mm} \caption{Upper panel: dependence of the energy gaps,
$\Delta\ped{\sigma}$ and $\Delta\ped{\pi}$, on the Al content $x$
in Mg$\ped{1-x}$Al$\ped{x}$B$_2$. The lines are only guides to the
eye. Lower panel: dependence of the gaps on the C content $y$ in
Mg(B$\ped{1-y}$C$\ped{y}$)$\ped{2}$. Lines are only guides to the
eye. }\label{fig:3}
\end{figure}

The gap values extracted from the fit of the conductance curves are
reported in Fig.\ref{fig:3} for Mg$_{1-x}$Al$_x$B$_2$ (upper panel)
and Mg(B$\ped{1-y}$C$\ped{y}$)$\ped{2}$ (lower panel). The trends
are roughly indicated by dashed curves that are simply guides to the
eye. The $x$ dependence of the gaps in Mg$_{1-x}$Al$_x$B$_2$ clearly
reflects the aforementioned distinction between two regimes,
delimited by the ``threshold'' value $x=0.09$. In the low-$x$
regime, the behaviour of $\Delta\ped{\sigma}$ and $\Delta\ped{\pi}$
is compatible with an increase in \emph{interband} scattering
\cite{Liu}. Within the two-band model in the Eliashberg formalism,
the gaps measured in the $x=0.08$ sample ($\Delta\ped{\pi}=3.1$ meV,
$\Delta\ped{\sigma}=6.1$ meV) can indeed be obtained from those of
MgB$_2$  by only increasing the interband scattering up to
$\Gamma\ped{\sigma \pi}\simeq 1.55$~meV. If one also takes into
account all the other effects of Al substitutions (i.e. the changes
in the DOS due to electron doping \cite{delaPena} and the stiffening
of the $E\ped{2g}$ phonon mode \cite{Postorino}), the experimental
trend of the gaps in the low-$x$ regime is qualitatively reproduced
\cite{Ummarino}. Things change completely above $x=0.09$, where the
experimental data contrast with all present theoretical models. For
high doping levels a merging of two gaps into one is predicted
either by simply increasing the interband scattering \cite{Liu} or
by including all the effects of Al substitution in the Eliashberg
theory~\cite{Ummarino}. In the first case, a BCS gap $\Delta \approx
4.1$~meV and a critical temperature $T\ped{c}\approx 26$~K are
expected in the so-called ``dirty'', isotropic limit~\cite{Liu}; in
the second case, the gap merging is predicted to occur around
$x=0.33$, when $\Delta\ped{\sigma}=\Delta\ped{\pi}\approx 3$~meV and
$T\ped{c}\simeq 20$~K \cite{Ummarino}. However, the experimental
results show no evidence of gap merging: $\Delta\ped{\pi}$ decreases
down to 0.4 meV at $x=0.21$ (with $T\ped{c}=20$~K) while
$\Delta\ped{\sigma}$ seems to saturate at about 4 meV. The failure
of the simple ``interband scattering'' picture is not really
surprising, since electron doping and phonon stiffening cannot be
neglected \cite{Postorino}. The failure of the more complete
Eliashberg two-band model is much more interesting. One possible
explanation is that phase segregation (indeed occurring in our
crystals at $x\gtrsim 0.10$) plays a major role in our result,
causing an unpredicted transition to a substantially different
physical system. However, it is worth saying that also recent gap
measurements in segregation-free Mg$\ped{1-x}$Al$\ped{x}$B$_2$
polycrystals by means of PCS \cite{Putti_nuovo} have given no
evidence of gap merging up to $x=0.3$, when $T\ped{c}$ is as low as
24 K. Instead, the values $\Delta\ped{\sigma}=2.0$~meV and
$\Delta\ped{\pi}=0.5$~meV have been found, that agree rather well
with those given by specific-heat measurements in the same samples
but contrast with the predictions of all present theories
\cite{Liu,Bussmann,Ummarino}.

The dependence of the gaps in Mg(B$\ped{1-y}$C$\ped{y}$)$\ped{2}$ on
the carbon content $y$, reported in the lower panel of
Fig.\ref{fig:3}, is much more regular. While $\Delta\ped{\pi}$
slightly increases, $\Delta\ped{\sigma}$ decreases monotonically
until, at $x=0.132$, \emph{only one gap} of amplitude $\Delta=3.2
\pm 0.9$ meV is observed \cite{nostro new}. Each point is the
average of different gap values measured in different contacts,
whose spread is indicated by the error bar. The large uncertainty at
$x=0.132$ may arise from carbon-content inhomogeneity on a length
scale of the order of $\xi$ \cite{Kazakov_C}, that unfortunately can
be detected by PCS. The overall gap trend does not differ much from
that predicted by the two-band Eliashberg model for Al
substitutions. Despite the different lattice sites occupied by Al
and C, some effects of the C substitution are indeed very similar to
those of Al doping: the decrease in $T\ped{c}$ (see
fig.\ref{fig:0}), the filling of the $\sigma$ bands due to electron
doping \cite{Singh}, the stiffening of the $E\ped{2g}$ phonon mode
and the consequent decrease in the electron-phonon coupling
\cite{Masui}. It is thus possible that, with suitable input from
experimental data, the two-band model can reproduce the results
presented here. As far as the ``interband scattering'' picture is
concerned, let us just remind that, in principle, carbon
substitutions should not increase the interband scattering
\cite{Erwin}. On the other hand, critical field measurements in
C-doped single crystals have evidenced a reduction in the
superconducting anisotropy~\cite{Masui}. The extrapolation of this
result above $y=0.10$ would lead to almost isotropic superconducting
properties accompanied by \emph{anisotropic} bandstructure, as would
be expected for strong interband scattering.

As we did in the case of Mg$_{1-x}$Al$_x$B$_2$, it is worth
comparing our results with other gap measurements in
Mg(B$\ped{1-y}$C$\ped{y}$)$\ped{2}$ reported in literature. Early
$\mu^{+}SR$ studies of Mg(B$\ped{1-y}$C$\ped{y}$)$\ped{2}$
polycrystals in the extreme low-doping region ($y\leq 0.03$)
\cite{Papagelis} showed a fast linear decrease of the gaps on
increasing $y$ (with the same slope for $\Delta\ped{\sigma}$ and
$\Delta\ped{\pi}$), in such a way that, at $y=0.03$,
$\Delta\ped{\sigma}=4.8$~meV and $\Delta\ped{\pi}=1.3$~meV. These
values are much smaller than ours, and also disagree with those
determined, in polycrystalline samples, by PCS \cite{Holanova} and
tunneling \cite{Schmidt}. In these last papers, the retention of
two-gap superconductivity was observed (as in our case) up to
$y=0.1$, where $T\ped{c}=22$~K. In the recent paper by
Ho\v{l}anov\'{a} \emph{et al.}, a linear decrease of the gaps vs
$T\ped{c}$ (with different slopes for $\Delta\ped{\sigma}$ and
$\Delta\ped{\pi}$) was claimed. The trend they evidenced for
$\Delta\ped{\sigma}$ is in very good agreement with ours, despite a
systematic difference in the absolute values (that are all smaller
than ours by 0.8 meV). On the contrary, the linear decrease in
$\Delta\ped{\pi}$ contrasts with our findings. These disagreements
might be due to the different nature and quality of the samples, but
further investigations are required to clarify this important point.

In conclusion, we have presented the results of the first systematic
investigation of Mg$_{1-x}$Al$_x$B$_2$ and
Mg(B$\ped{1-y}$C$\ped{y}$)$\ped{2}$ single crystals by directional
point-contact spectroscopy. We have shown that the dependence of the
gaps on the Al content contrasts will all present theories and is
probably affected by phase segregation above $x\simeq 0.10$. In
C-substituted crystals, instead, we have found the first evidence of
gap merging, predicted theoretically as a result of the
doping-induced changes in the DOS and in the phonon frequency and/or
by the increase of interband scattering. This finding might provide
the longed for, final test of the theoretical models for two-band
superconductivity.

This work was done within the Project PRA ``UMBRA'' of INFM and
the INTAS Project no. 01-0617. V.A.S. acknowledges the support
from RFBR (project no. 04-02-1726) and the Ministry of Science and
Technologies of the Russian Federation (contract no.
40.012.1.1.1357).

\end{document}